\documentclass[journal]{IEEEtran}
\usepackage{cite}
\usepackage{amsmath,amssymb,amsfonts}
\usepackage{graphicx}
\usepackage{multirow}
\usepackage{placeins}
\usepackage{float}
\usepackage{textcomp}
\usepackage{hyperref}
\hypersetup{
  hidelinks=true,
  pdftitle={EEG-based Visual Retrieval and Reconstruction: From Neurally Visible Optimal Layer to Hierarchical Diffusion Generation},
  pdfauthor={Minyi Wang, Zhenqin Wu, Rihui Li}
}
\graphicspath{{./figures/}}
\newcommand{\figonesubfig}[2]{%
  \par\noindent
  \begin{minipage}{\columnwidth}
    \raggedright
    {\footnotesize\mdseries(#1)}\\[0.12em]
    \includegraphics[width=\columnwidth]{#2}%
  \end{minipage}%
  \par\vspace{0.1em}%
}
\newsavebox{\figcornerbox}
\newlength{\figfivew}

\def\BibTeX{{\rm B\kern-.05em{\sc i\kern-.025em b}\kern-.08em
    T\kern-.1667em\lower.7ex\hbox{E}\kern-.125emX}}
\begin{document}
\title{EEG-based Visual Retrieval and Reconstruction: From Neurally Visible Optimal Layer to Hierarchical Diffusion Generation}
\author{Minyi Wang, Zhenqin Wu and Rihui Li
\thanks{This work was supported by the Science and Technology Development Fund of the Macao SAR (0016/2024/RIB1), Guangdong Provincial Natural Science Foundation (2025A1515010539), and the University of Macau (MYRG-GRG2024-00296-ICI, and MYRG-CRG2024-00022-ICI).(Corresponding author: Rihui Li.)}
\thanks{Minyi Wang is with the Centre for Cognitive and Brain Sciences, Institute of Artificial Intelligence and Brain Sciences, University of Macau, Taipa, Macau S.A.R. 999078, China (e-mail: minyiwang11@gmail.com).}
\thanks{Zhenqin Wu is with the School of Biomedical Engineering and the School of Computing and Data Science, The University of Hong Kong, Hong Kong SAR (e-mail: zqwu@cs.hku.hk).}
\thanks{Rihui Li is with the Centre for Cognitive and Brain Sciences, Institute of Artificial Intelligence and Brain Sciences, University of Macau, Taipa, Macau S.A.R. 999078, China, and also with the Department of Biomedical Engineering, Faculty of Engineering, University of Macau, Taipa, Macau S.A.R. 999078, China (e-mail: rihuili@um.edu.mo).}
}

\maketitle

\begin{abstract}
Decoding visual perception from electroencephalography (EEG) is important for non-invasive brain-computer interfaces (BCIs). However, most existing visual decoding pipelines directly align EEG features with semantic features from pretrained vision models. Those EEG signals carry information at more than one level and this practice disregards the varying neural visibility of different visual components in EEG signals, leading to cross modal mismatches and incomplete information use. In this work, we address this limitation through layer-wise contrastive learning. For each subject, the intermediate CLIP layer that maximizes retrieval performance is selected as the Neural Visibility Optimal Layer (NVOL). Built on NVOL, a hierarchical framework couples retrieval and generation through a shared intermediate representation. The retrieval branch fuses multi-NVOL features, aligns them to image embeddings via contrastive learning, and applies cross-domain similarity local scaling (CSLS) at test time to mitigate hubness. The generation branch reconstructs subject-specific NVOL features from EEG using a conditional diffusion prior, maps them to CLIP space through a lightweight adapter, and drives a pretrained Stable Diffusion XL model. Experimental validation on THINGS-EEG showed that, NVOL-based retrieval achieves 78.1\% mean Top-1 accuracy in 200-way retrieval, rising to 86.4\% with CSLS. Two-stage NVOL-to-semantic reconstruction also outperforms single-stage final-layer diffusion on semantic and structural metrics. By aligning EEG with layer-wise neural visibility rather than fixed high-level semantics, the proposed framework improves both retrieval accuracy and image reconstruction in EEG-based visual decoding.

\end{abstract}


\section{Introduction}
\label{sec:introduction}

Visual perception supports humans' recognition and interaction with the external environment. A central question in neuroscience and brain--computer interface (BCI) research is whether visual neural signal can be decoded into machine-readable representations \cite{ref1,ref2}. With the development of visual-language models \cite{ref3,ref4}, researchers can build human-computer interaction systems that can directly read and understand the visual signals of the brain. These systems may clarify how the brain represents external visual input and may also support practical BCI applications outside the laboratory.

Visual decoding first requires appropriate neural recordings. Although functional magnetic resonance imaging (fMRI) provides high spatial resolution and has long been a primary tool for visual decoding \cite{ref5}, it is costly, temporally limited, and difficult to deploy in portable or real-time settings \cite{ref6}. Electroencephalography (EEG) offers higher temporal resolution, portability, and lower cost \cite{ref7}, although scalp recordings are noisy and spatially coarse, so extracting stable visual representations from EEG remains a core methodological challenge \cite{ref8}.

EEG-based visual decoding has moved beyond image classification to object-level decoding and image reconstruction \cite{ref9,ref10}. Large EEG datasets combined with deep learning have improved decoding accuracy \cite{ref11}. Diffusion-based priors have reconstructed natural images from fMRI and magnetoencephalography(MEG) \cite{ref12,ref13,ref14}, and similar paradigms are now being adapted to EEG despite its lower spatial resolution and higher noise. However, as reconstruction becomes a primary goal \cite{ref10}, it remains poorly defined how visual information is organized across representational levels in scalp EEG and which levels support reliable alignment with computational vision models \cite{ref10,ref15}.

Prior studies suggested that EEG does not encode visual features uniformly across the hierarchy \cite{ref16}. Global structure linked to low-frequency components is relatively stable in scalp recordings, whereas high-level semantic information shows lower neural visibility \cite{ref15}. Most EEG visual decoding methods still follow a vision--language style pipeline: they use contrastive learning to align EEG features with image features from a pretrained vision model, and they take the final-layer semantic embedding as the target \cite{ref17}. This ignores the hierarchical nature of visual processing in the brain. The resulting highly compressed features may match poorly with the multi-scale information available in scalp EEG. In EEG2Vision, reducing the number of electrodes dropped category decoding accuracy from 89\% to 38\%, while reconstruction quality changed only modestly \cite{ref18}. In structure-guided diffusion models, explicit structural cues improved both low-level visual attributes and semantic content \cite{ref19}.

Recent work has added structural guidance \cite{ref19}, built staged coarse- and fine-grained semantic representations \cite{ref20}, or used two-stage diffusion pipelines that keep low-level detail but weaken semantic coherence. Staged representation learning that separates coarse- and fine-grained semantics reached 84.6\% zero-shot decoding accuracy on THINGS-EEG, 21.4\% points above the baseline \cite{ref15,ref20}. EEG retrieval systems trained with InfoNCE-style objectives can improve cross-modal alignment \cite{ref21}, but high-dimensional embeddings often show hubness, where a few classes dominate nearest-neighbor lists \cite{ref22}. Generation methods also tend to reuse EEG features learned for retrieval; when retrieval is unstable, reconstruction can suffer as well. These observations suggest that EEG decoding needs visual layers that are more neurally visible, retrieval calibration that reduces hubness, and a shared hierarchical representation for both retrieval and generation \cite{ref15,ref20,ref22,ref23}.

In this study, we built a unified hierarchical alignment framework that places retrieval and generation in a shared feature space. Instead of the final semantic layer, we aligned EEG with intermediate layers of the vision encoder. Our main contributions are as follows.

\begin{itemize}

\item We examined EEG--image alignment from shallow to deep layers and found that the best alignment depth varied across participants, but remained in intermediate layers.

\item We fused multiple neurally visible layers rather than forcing a single alignment target, with the aim of improving representational capacity and interpretability.

\item We built a unified retrieval--generation pipeline (EEG $\rightarrow$ neurally visible optimal layer $\rightarrow$ final layer). The retrieval branch trained an EEG encoder with contrastive learning and applied CSLS at test time to reduce hubness. The generation branch reconstructed neurally visible features with a diffusion prior and mapped them to the final CLIP layer through a lightweight adapter for SDXL image generation.

\end{itemize}

\section{Related Work}
\label{sec:relatedwork}

\subsection{Neural-signal visual decoding}

Visual decoding from non-invasive brain signals has moved from traditional classifiers to end-to-end deep models. Early EEG work used linear decoders and handcrafted features; later studies adopted compact CNNs such as EEGNet to model spatiotemporal pattern \cite{ref24}. Large-scale benchmarks, including THINGS-EEG, have enabled training of systems for semantic retrieval and image generation \cite{ref1,ref11}. Diffusion-based priors have also been used to reconstruct natural images from fMRI and MEG, and similar approaches are now being adapted to EEG despite its lower spatial resolution and higher noise \cite{ref12,ref13,ref14}.

\subsection{Visual foundation models and brain-feature alignment}

Foundation vision encoders such as ViT and CLIP provide hierarchical representations from low-level structure to high-level semantics \cite{ref3,ref25}. Many brain-alignment methods map neural signals to the final semantic layer, which may suffice for coarse category prediction but can under-use intermediate layers that are more neurally observable \cite{ref26}. Contrastive alignment has improved EEG retrieval, but most systems still use one fixed target layer for all subjects.

\subsection{Contrastive Learning for Brain Decoding}

Cross-modal retrieval is often trained with InfoNCE-style objectives \cite{ref21}, but high-dimensional embedding spaces can show hubness, in which a few classes dominate nearest-neighbor lists \cite{ref22}. CSLS was proposed to reduce this bias in bilingual retrieval and is now used as a post-hoc calibration step in cross-domain matching \cite{ref23}. For generation, DDPM and latent diffusion models provide stable, high-fidelity synthesis \cite{ref27}. Our method combines intermediate-layer fusion with contrastive retrieval alignment, CSLS for hubness mitigation, and a two-stage diffusion pipeline for reconstruction.

\section{Method}
\label{sec:method}

\subsection{Overview}

\begin{figure*}[!t]
\centering
\includegraphics[width=\textwidth]{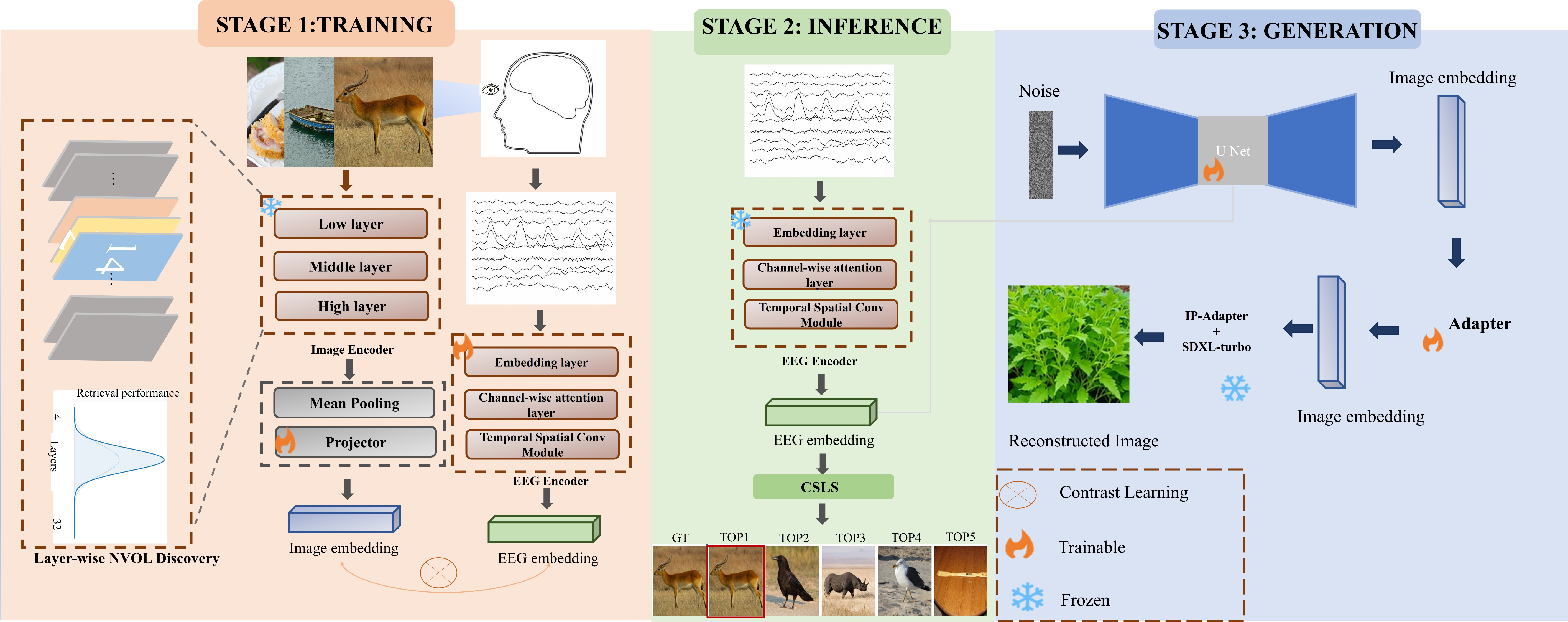}
\caption{Framework of EEG-based visual decoding and generation. Training stage: The image encoder extracts image embeddings, and the EEG encoder extracts EEG embeddings; matched image-EEG pairs are pulled closer via contrastive learning. Inference stage: The trained EEG encoder maps the test EEG to embeddings, and zero-shot retrieval is achieved with CSLS post-processing. Generation stage: The EEG embeddings learned during the retrieval stage serve as conditional inputs to a U-Net to decode image-space embeddings, which are then passed through an Adapter and fed into SDXL for image reconstruction}
\label{fig1}
\end{figure*}

We proposed a hierarchical framework for EEG-based visual retrieval and reconstruction (Fig.~\ref{fig1}). Given a raw EEG recording, the pipeline aligned EEG embeddings with hierarchical semantic features from CLIP intermediate layers for retrieval, then used those intermediate representations as a bridge for image reconstruction.

\subsection{Retrieval Branch}

\subsubsection{EEG Encoder}

\begin{figure}[!t]
\centering
\includegraphics[width=\columnwidth]{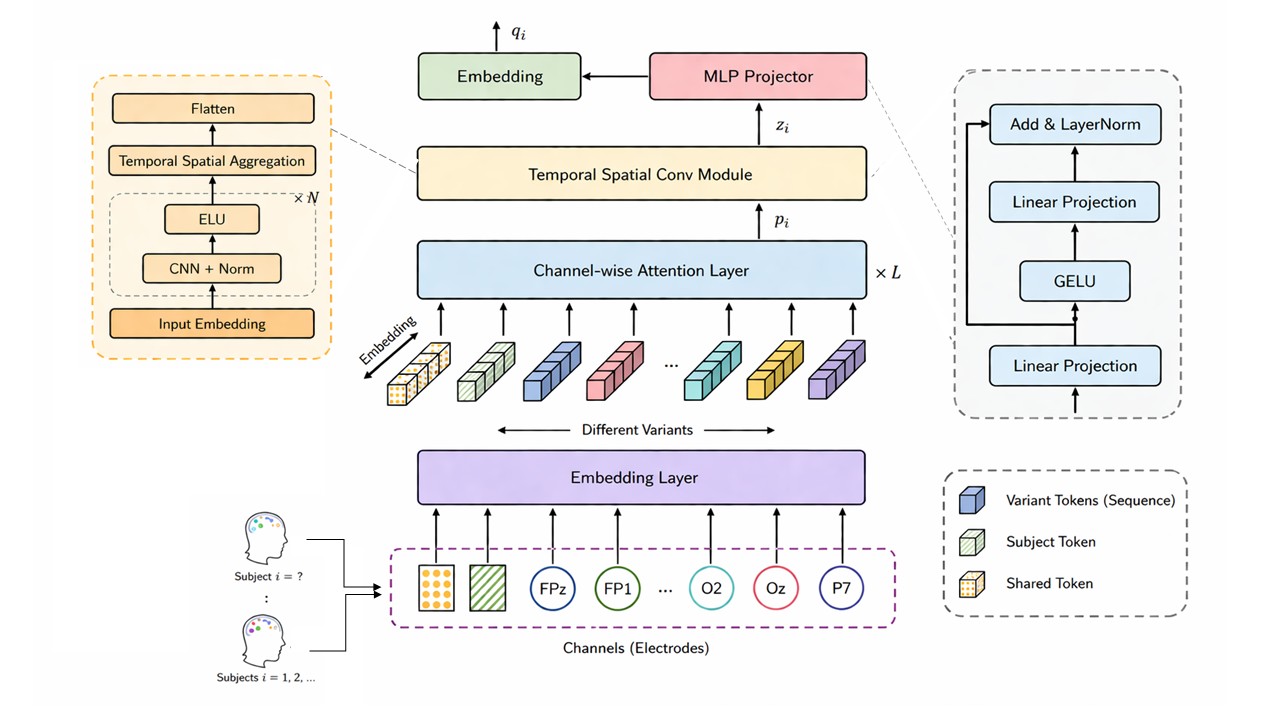}
\caption{Architecture of the Adaptive Thinking Mapper (ATM) EEG encoder \cite{ref28}.}
\label{fig2}
\end{figure}

We used the Adaptive Thinking Mapper (ATM) architecture to extract neural embeddings \cite{ref28}. The encoder applied a temporal--spatial convolution block to capture local waveform patterns, followed by a Transformer-based global integration layer. The final output was a normalized 1024-dimensional embedding $e\in \mathbb{R}^{1024}$. The structure of the ATM encoder was illustrated in Fig.~\ref{fig2}.

\subsubsection{NVOL Selection and Multi-layer Fusion}

We extracted features from a pretrained CLIP vision encoder (ViT-H/14) comprising 32 Transformer blocks with hidden dimension $d=1280$. For an input image $I$, let $f_l(I)\in\mathbb{R}^{d\times N_t}$ denote the token-wise features from block $l$, where $N_t$ is the number of visual tokens. Layer-wise retrieval was screened across layers $L=\{4,6,8,10,12,14,16,18,20,22,24,26,28,30, 32\}$ and peak alignment occurred in intermediate layers (Table~\ref{tab1}). We therefore selected $L=\{10,12,14\}$ as neural-visible optimal layers (NVOLs) for fusion.

For each selected layer, mean pooling over the token dimension yielded a global descriptor:
\begin{equation}
{h}_{l}=\left(1/{N}_{t}\right)\sum_{i=1}^{N_t}{f}_{l,i}\left(I\right),\quad {h}_{l}\in \mathbb{R}^{d}
\label{eq1}
\end{equation}
Each ${h}_{l}$ was then mapped to a shared latent dimension $D=1024$ by a layer-specific linear projection followed by layer normalization:
\begin{equation}
{u}_{l}=\mathrm{LayerNorm}\left({W}_{l}{h}_{l}+{b}_{l}\right),\quad {u}_{l}\in \mathbb{R}^{D}
\label{eq2}
\end{equation}

Projected features from the selected layers were fused with learnable weights $w = (w_1,\ldots,w_{|L|})^\top$:
\begin{gather}
{v}_{raw}=\sum_{l\in L}\mathrm{softmax}\left(w\right)_{l}\cdot {u}_{l}
\label{eq3}\\
v={v}_{raw}/{\|{v}_{raw}\|}_{2}
\label{eq4}
\end{gather}
where $\mathrm{softmax}(w)$ assigns non-negative, normalized weights across layers. The fused embedding $v$ was used as the visual target for contrastive alignment and as the NVOL supervision target in the generation branch.

\subsubsection{Contrastive Training and CSLS Retrieval}

Let $E_{\theta}$ denote the EEG encoder and $F_{\phi}$ the fused visual encoder. Given EEG input $x$ and image $I$, the embeddings $e=E_{\theta}(x)$ and $v=F_{\phi}(I)$ were L2-normalized to $\hat{e}=e/\|e\|_2$ and $\hat{v}=v/\|v\|_2$. During training, $e$ and $v$ were aligned with a symmetric InfoNCE loss:
\begin{equation}
\begin{split}
{L}_{InfoNCE}=&-\frac{1}{2N}\sum_{i=1}\Big(\log \frac{\exp \left({e}_{i}\cdot {v}_{i}/\tau\right)}{\sum_{j=1}\exp \left({e}_{i}\cdot {v}_{j}/\tau\right)}\\
&+\log \frac{\exp \left({v}_{i}\cdot {e}_{i}/\tau\right)}{\sum_{j=1}\exp \left({v}_{i}\cdot {e}_{j}/\tau\right)}\Big)
\end{split}
\label{eq6}
\end{equation}
where $\tau$ is a learnable log-temperature. At inference, cross-modal similarity was scored as a temperature-scaled dot product
\begin{equation}
s\left(x,I\right)=\hat{e}^{\top}\hat{v}
\label{eq7}
\end{equation}

Each EEG query was evaluated in an $N$-way zero-shot classification task \cite{ref29}. Top-1 and Top-5 accuracies were obtained from the ranked similarity scores over the $N$ held-out candidate classes and averaged across subjects. Category-level hubness was quantified from $k$-NN retrieval statistics \cite{ref30}. Let $N_k(c)$ denote how often category $c$ appears in the top-$k$ neighbors across queries, computed separately for image-to-image and EEG-to-image retrieval. Skew was summarized by the Gini coefficient \cite{ref31}:
\begin{equation}
G=\frac{\sum_{i=1}^{C}\left(2i-C-1\right)\cdot N_{k}(i)}{C\cdot \sum_{c}{N}_{k}\left(c\right)}
\label{eq8}
\end{equation}
where $C$ is the number of categories and $N_k(i)$ is the $i$-th smallest hub count. Values near $0$ indicate near-uniform retrieval frequency, whereas values approaching $1$ indicate strong hubness.

\begin{table*}[!t]
	\caption{Zero-shot retrieval performance on the Things-EEG dataset. Top-1 and Top-5 accuracy (\%) are reported for each subject and the average, broken down by layers of the CLIP model.}
	\label{tab1}
	\centering
	\tiny
	\resizebox{0.75\textwidth}{!}{%
		\begin{tabular}{llccccccccccc}
			\hline
			Layers &  & Sub1 & Sub2 & Sub3 & Sub4 & Sub5 & Sub6 & Sub7 & Sub8 & Sub9 & Sub10 & Avg \\
			\hline
			Layer 4 & Top1 & 68.5 & 56.0 & 56.5 & 45.0 & 56.5 & 67.5 & 61.0 & 67.0 & 57.0 & 53.5 & 58.9 \\
			Layer 4 & Top5 & 94.5 & 90.5 & 86.5 & 85.0 & 85.5 & 94.0 & 89.5 & 93.5 & 83.5 & 85.0 & 88.7 \\
			Layer 6 & Top1 & 67.5 & 61.0 & 63.5 & 51.5 & 56.5 & 69.5 & 63.0 & 67.0 & 58.0 & 59.5 & 61.7 \\
			Layer 6 & Top5 & 92.5 & 92.0 & 90.5 & 87.5 & 84.0 & 94.0 & 90.5 & 92.5 & 86.5 & 87.5 & 89.8 \\
			Layer 8 & Top1 & 72.0 & 67.0 & 65.0 & 58.5 & 55.5 & 77.5 & 68.0 & 78.5 & 57.0 & 70.5 & 67.0 \\
			Layer 8 & Top5 & 95.0 & 90.0 & 89.0 & 86.5 & 82.0 & 96.0 & 95.5 & 96.0 & 89.5 & 92.0 & 91.1 \\
			Layer 10 & Top1 & 75.0 & 72.0 & 67.5 & 67.0 & 62.0 & 79.0 & 66.5 & 76.5 & 63.5 & 68.5 & 69.8 \\
			Layer 10 & Top5 & 97.5 & 93.5 & 94.5 & 92.0 & 85.5 & 96.5 & 95.5 & 98.5 & 89.5 & 93.0 & 93.6 \\
			Layer 12 & Top1 & 74.5 & 72.0 & 76.5 & 68.0 & 66.5 & 74.5 & 73.0 & 79.5 & 70.5 & 72.5 & 72.8 \\
			Layer 12 & Top5 & 97.0 & 94.0 & 94.5 & 93.0 & 83.5 & 96.5 & 93.0 & 98.0 & 93.5 & 96.0 & 93.9 \\
			Layer 14 & Top1 & 70.5 & 74.0 & 69.0 & 69.5 & 61.0 & 73.0 & 70.0 & 76.5 & 64.5 & 69.0 & 69.7 \\
			Layer 14 & Top5 & 96.5 & 94.0 & 93.0 & 91.0 & 88.0 & 94.0 & 95.0 & 96.0 & 92.5 & 94.5 & 93.5 \\
			Layer 16 & Top1 & 64.5 & 59.0 & 64.0 & 66.0 & 51.5 & 70.5 & 69.5 & 70.5 & 61.0 & 67.0 & 64.3 \\
			Layer 16 & Top5 & 93.0 & 90.5 & 93.0 & 90.5 & 80.5 & 91.0 & 93.0 & 94.5 & 91.0 & 93.0 & 91.0 \\
			Layer 18 & Top1 & 61.5 & 52.0 & 58.5 & 61.0 & 49.0 & 62.0 & 60.5 & 71.0 & 51.0 & 62.0 & 58.9 \\
			Layer 18 & Top5 & 88.5 & 87.5 & 88.5 & 89.5 & 83.5 & 90.0 & 88.0 & 90.0 & 85.0 & 90.0 & 88.1 \\
			Layer 20 & Top1 & 56.5 & 47.5 & 55.0 & 52.5 & 46.5 & 50.0 & 53.0 & 67.0 & 52.5 & 54.5 & 53.5 \\
			Layer 20 & Top5 & 86.0 & 82.5 & 84.0 & 84.0 & 75.5 & 82.5 & 86.0 & 90.0 & 84.0 & 89.0 & 84.4 \\
			Layer 22 & Top1 & 45.0 & 41.0 & 48.0 & 44.5 & 39.5 & 47.5 & 50.0 & 60.0 & 48.5 & 49.5 & 47.3 \\
			Layer 22 & Top5 & 79.5 & 76.5 & 79.5 & 80.0 & 68.0 & 78.5 & 77.5 & 85.0 & 78.0 & 80.5 & 78.3 \\
			Layer 24 & Top1 & 43.5 & 41.0 & 43.5 & 48.0 & 36.5 & 46.5 & 40.5 & 57.0 & 43.5 & 41.5 & 44.1 \\
			Layer 24 & Top5 & 74.0 & 67.5 & 77.5 & 78.5 & 69.5 & 78.0 & 71.0 & 85.5 & 72.5 & 78.5 & 75.2 \\
			Layer 26 & Top1 & 37.5 & 41.0 & 41.5 & 42.5 & 32.0 & 42.0 & 40.5 & 51.0 & 37.0 & 39.0 & 40.4 \\
			Layer 26 & Top5 & 73.5 & 68.5 & 71.0 & 78.5 & 60.0 & 73.5 & 72.0 & 83.5 & 70.0 & 76.5 & 72.7 \\
			Layer 28 & Top1 & 39.0 & 38.0 & 41.5 & 37.0 & 28.0 & 40.0 & 37.5 & 53.5 & 38.0 & 37.5 & 39.0 \\
			Layer 28 & Top5 & 69.5 & 67.0 & 73.0 & 69.0 & 56.0 & 73.5 & 69.0 & 84.5 & 67.5 & 72.0 & 70.1 \\
			Layer 30 & Top1 & 32.0 & 25.0 & 33.0 & 31.0 & 27.0 & 34.0 & 35.5 & 44.5 & 33.5 & 37.5 & 33.3 \\
			Layer 30 & Top5 & 63.5 & 58.5 & 66.0 & 70.5 & 53.5 & 65.0 & 68.5 & 82.0 & 66.5 & 68.0 & 66.2 \\
			Layer 32 & Top1 & 21.0 & 25.0 & 30.0 & 30.0 & 22.0 & 28.0 & 26.0 & 41.5 & 30.0 & 33.0 & 28.6 \\
			Layer 32 & Top5 & 53.0 & 57.0 & 63.0 & 61.0 & 49.5 & 58.0 & 58.5 & 76.0 & 59.5 & 67.0 & 60.2 \\
			\hline
		\end{tabular}%
	}
\end{table*}

We applied Cross-domain Similarity Local Scaling (CSLS) \cite{ref22,ref32} to calibrate raw similarities for EEG-to-image retrieval. The adjusted score between EEG query $x$ and image category $c$ is
\begin{gather}
{s}_{CSLS}\left(x,c\right)=2s\left(x,c\right)-{r}_{E}\left(x\right)-{r}_{I}\left(c\right),
\label{eq9}\\
{r}_{E}\left(x\right)=\frac{1}{k}\sum_{{I}_{j}\in {N}_{k}\left(x\right)}s\left(x,{I}_{j}\right),
\label{eq10}\\
{r}_{I}\left(c\right)=\frac{1}{k}\sum_{{x}_{j}\in {N}_{k}\left(c\right)}s\left({x}_{j},c\right),
\label{eq11}
\end{gather}
where ${N}_{k}\left(x\right)$ and ${N}_{k}\left(c\right)$ denote the $k$ nearest image neighbors of $x$ and the $k$ nearest EEG queries of category $c$, respectively. We first constructed a global similarity matrix $S$ where $S_{ij}=s(x_i,I_j)$, then computed row-wise and column-wise top-$k$ averages to obtain ${r}_{E}$ and ${r}_{I}$, respectively.

To determine whether the observed hubness exceeded that expected from finite-sample random retrieval, we constructed a random-retrieval null model separately for each $k$. Under the null hypothesis, each of $Q=200$ queries selected $k$ distinct categories uniformly without replacement from a gallery of $C=200$ categories. We generated $20{,}000$ single-subject null Gini coefficients. To match the group-level analysis, $B=100{,}000$ null group means were then generated by sampling $10$ single-subject null values and averaging them. The significance threshold was defined as the 95th percentile of the corresponding group-mean null distribution:
\begin{equation}
T_{k,\alpha}=F_{0,k}^{-1}\left(1-\alpha\right),\quad \alpha=0.05,
\label{eq12}
\end{equation}
where $F_{0,k}$ denotes the null distribution of the group-mean Gini coefficient for neighborhood size $k$. One-sided Monte Carlo $p$-values were calculated as
\begin{equation}
p=\frac{1+\sum_{b=1}^{B}\mathbb{I}\left(G_{0,k}^{\left(b\right)}\geq G_{\mathrm{obs},k}\right)}{B+1},
\label{eq13}
\end{equation}
where $G_{\mathrm{obs},k}$ is the observed group-mean Gini coefficient and $G_{0,k}^{\left(b\right)}$ is the $b$-th null replicate. Because hubness was tested across $15$ layers and two neighborhood sizes, the resulting $30$ $p$-values were corrected using the Benjamini--Hochberg false discovery rate procedure.

\subsection{Generation Branch}

Direct mapping from noisy EEG to final CLIP semantic space can lead to representation collapse. We therefore used a two-stage generative pipeline: (1) an EEG-to-NVOL diffusion prior that reconstructed intermediate CLIP features, and (2) an NVOL-to-semantic adapter that maped reconstructed NVOL to final-layer CLIP embeddings for SDXL conditioning.

\begin{table*}[!t]
	\caption{Zero-shot retrieval performance on the THINGS-EEG dataset. The table reports the Top-1 and Top-5 accuracies (\%) for the subjects and their average. Bold values indicate the best performance.}
	\label{tab2}
	\centering
	\tiny
	\resizebox{0.75\textwidth}{!}{%
		\begin{tabular}{llccccccccccc}
			\hline
			Method &  & Sub1 & Sub2 & Sub3 & Sub4 & Sub5 & Sub6 & Sub7 & Sub8 & Sub9 & Sub10 & Avg \\
			\hline
			\multicolumn{13}{c}{\textit{Intra-subject: train and test on one subject}} \\
			\hline
			\multirow{2}{*}{BraVL~\cite{ref50}} & Top1 & 6.1 & 4.9 & 5.6 & 5.0 & 4.0 & 6.0 & 6.5 & 8.8 & 4.3 & 7.0 & 5.8 \\
			 & Top5 & 17.9 & 14.9 & 17.4 & 15.1 & 13.4 & 18.2 & 20.4 & 23.7 & 14.0 & 19.7 & 17.5 \\
			\multirow{2}{*}{NICE~\cite{ref29}} & Top1 & 13.2 & 12.5 & 14.5 & 15.3 & 10.1 & 16.5 & 17.0 & 12.9 & 15.4 & 17.4 & 14.1 \\
			 & Top5 & 39.5 & 40.3 & 42.7 & 39.6 & 31.5 & 44.0 & 42.1 & 56.1 & 41.6 & 45.8 & 43.6 \\
			\multirow{2}{*}{ATM~\cite{ref28}} & Top1 & 25.6 & 22.0 & 25.0 & 31.4 & 12.9 & 21.3 & 30.5 & 38.8 & 34.4 & 29.1 & 27.1 \\
			 & Top5 & 60.4 & 54.5 & 62.4 & 60.9 & 43.0 & 51.1 & 61.5 & 72.0 & 51.5 & 63.5 & 58.4 \\
			\multirow{2}{*}{CogCap~\cite{ref51}} & Top1 & 27.2 & 28.7 & 37.2 & 27.7 & 21.8 & 31.6 & 32.8 & 47.6 & 33.4 & 35.1 & 33.3 \\
			 & Top5 & 59.5 & 57.0 & 66.1 & 63.2 & 47.8 & 58.1 & 59.6 & 73.5 & 57.7 & 63.6 & 60.6 \\
			\multirow{2}{*}{VE-SDN~\cite{ref52}} & Top1 & 32.6 & 34.4 & 38.7 & 39.8 & 29.4 & 34.5 & 34.5 & 49.3 & 39.0 & 39.8 & 37.2 \\
			 & Top5 & 63.7 & 69.9 & 73.5 & 72.0 & 58.6 & 68.8 & 68.3 & 79.8 & 69.6 & 75.3 & 70.0 \\
			\multirow{2}{*}{UBP~\cite{ref53}} & Top1 & 41.2 & 51.2 & 51.2 & 51.1 & 42.2 & 57.5 & 49.0 & 58.6 & 45.1 & 61.5 & 50.9 \\
			 & Top5 & 70.5 & 80.9 & 82.0 & 76.9 & 72.8 & 83.5 & 83.5 & 85.8 & 76.2 & 88.2 & 79.7 \\
			\multirow{2}{*}{Ours(single layer)} & Top1 & 75.0 & 74.0 & 76.5 & 69.5 & 66.5 & 79.0 & 73.0 & 79.5 & 70.5 & 72.5 & 73.6 \\
			 & Top5 & 97.5 & 94.0 & 94.5 & 91.0 & 83.5 & 96.5 & 93.0 & 98.0 & 93.5 & 96.0 & 93.8 \\
			\multirow{2}{*}{Ours(single layer + CSLS)} & Top1 & 87.5 & 84.0 & 90.0 & 80.5 & 73.0 & 89.0 & 84.5 & 90.0 & 81.5 & 85.5 & 84.6 \\
			 & Top5 & 99.5 & 97.5 & 99.0 & 94.5 & 94.5 & 98.5 & 98.5 & 99.5 & 98.5 & 96.5 & 97.7 \\
			\multirow{2}{*}{Ours(multilayer fusion)} & Top1 & 85.0 & 82.5 & 77.5 & 73.0 & 72.0 & 77.5 & 78.5 & 84.5 & 77.5 & 72.5 & 78.1 \\
			 & Top5 & 98.0 & 97.5 & 97.5 & 96.0 & 95.0 & 96.5 & 96.0 & 98.0 & 95.5 & 96.5 & 96.7 \\
			\multirow{2}{*}{Ours(multilayer + CSLS)} & Top1 & 92.0 & 91.5 & 84.5 & 81.0 & 80.0 & 88.0 & 82.5 & 91.0 & 86.5 & 87.0 & \textbf{86.4} \\
			 & Top5 & 100 & 98.5 & 100.0 & 97.0 & 98.0 & 98.0 & 97.0 & 99.5 & 98.5 & 98.0 & \textbf{98.5} \\
			\hline
			\multicolumn{13}{c}{\textit{Inter-Subject: leave one subject out for test}} \\
			\hline
			\multirow{2}{*}{BraVL~\cite{ref50}} & Top1 & 2.3 & 1.5 & 1.4 & 1.7 & 1.5 & 1.8 & 2.1 & 2.2 & 1.6 & 2.3 & 1.8 \\
			 & Top5 & 8.0 & 6.3 & 5.9 & 6.7 & 5.6 & 7.2 & 8.1 & 7.6 & 6.4 & 8.5 & 7.0 \\
			\multirow{2}{*}{NICE~\cite{ref29}} & Top1 & 7.6 & 5.9 & 6.0 & 6.3 & 4.4 & 5.6 & 6.3 & 5.7 & 8.4 & 6.2 & 6.2 \\
			 & Top5 & 22.8 & 20.5 & 22.3 & 20.7 & 18.3 & 22.2 & 19.7 & 22.0 & 17.6 & 28.3 & 21.4 \\
			\multirow{2}{*}{ATM~\cite{ref28}} & Top1 & 10.5 & 7.1 & 11.9 & 14.7 & 7.0 & 11.1 & 16.1 & 15.0 & 4.9 & 20.5 & 11.9 \\
			 & Top5 & 26.8 & 24.8 & 33.8 & 39.4 & 23.9 & 35.8 & 43.5 & 40.3 & 22.7 & 46.5 & 33.8 \\
			\multirow{2}{*}{UBP~\cite{ref53}} & Top1 & 11.5 & 15.5 & 9.8 & 13.0 & 8.8 & 11.7 & 10.2 & 12.2 & 15.5 & 16.0 & 12.4 \\
			 & Top5 & 29.7 & 40.0 & 27.0 & 32.3 & 33.8 & 31.0 & 23.8 & 32.2 & 40.5 & 43.5 & 33.4 \\
			\multirow{2}{*}{Ours(multilayer fusion)} & Top1 & 17.0 & 11.0 & 6.5 & 14.5 & 5.5 & 17.0 & 15.5 & 10.5 & 5.5 & 23.5 & 12.7 \\
			 & Top5 & 41.0 & 27.0 & 15.0 & 30.5 & 15.5 & 34.5 & 41.0 & 27.0 & 13.5 & 47.0 & 29.2 \\
			\multirow{2}{*}{Ours(multilayer + CSLS)} & Top1 & 24.5 & 21.0 & 8.0 & 22.0 & 11.5 & 24.5 & 25.0 & 17.0 & 9.0 & 36.0 & \textbf{19.9} \\
			 & Top5 & 60.0 & 50.5 & 20.5 & 44.0 & 30.5 & 53.5 & 57.0 & 37.5 & 22.5 & 64.5 & \textbf{44.1} \\
			\hline
		\end{tabular}%
	}
\end{table*}

\subsubsection{EEG-to-NVOL Diffusion Prior}

Given EEG embedding $e$, we modeled the conditional distribution of intermediate CLIP features $v$ with a denoising diffusion probabilistic model (DDPM) \cite{ref27}. Let $v_t$ denote the noisy latent at diffusion step $t$, and let the denoiser $G_{\psi}$ predict the noise $\epsilon$ added to $v$:
\begin{equation}
\epsilon = G_{\psi}(v_t, t, e).
\end{equation}
The standard noise-prediction loss was
\begin{equation}
\mathcal{L}_{\mathrm{noise}} = \mathbb{E}_{t,\epsilon}\left[\left\|\epsilon - G_{\psi}(v_t, t, e)\right\|_2^2\right]
\end{equation}
Noise prediction alone did not fully constrain the denoised latent to the true NVOL manifold. We also reconstructed the clean latent $\hat{v}_0$ from $\epsilon$ using the DDPM identity and applied explicit supervision in Euclidean and angular space:
\begin{gather}
\mathcal{L}_{\mathrm{MSE}} = \left\|\hat{v}_0 - v\right\|_2^2
\label{eq14}\\
\mathcal{L}_{\mathrm{cos}} = 1 - \cos\left(\hat{v}_0, v\right)
\label{eq15}
\end{gather}
where $\cos(\cdot,\cdot)$ is computed after L2 normalization.

We also added a pairwise geometry loss to match batch-wise relational structure between predicted and target NVOL embeddings. This term preserves similarity patterns among semantically related and unrelated samples within each mini-batch and is related to similarity-keeping regularization in EEG--image contrastive learning \cite{ref33,ref34,ref35}. For a mini-batch of size $B$, let $\hat{V}, V \in \mathbb{R}^{B \times D}$ denote row-normalized predicted and ground-truth NVOL embeddings. We computed cosine-similarity (Gram) matrices
\begin{equation}
M_{\mathrm{pred}} = \hat{V}\hat{V}^{\top}, \quad M_{\mathrm{gt}} = V V^{\top}
\label{eq16}
\end{equation}
and defined the geometry loss as the normalized squared Frobenius norm of their difference:
\begin{equation}
\mathcal{L}_{\mathrm{geo}} = \frac{1}{B^2}\left\|M_{\mathrm{pred}} - M_{\mathrm{gt}}\right\|_F^2
\label{eq17}
\end{equation}
Minimizing $\mathcal{L}_{\mathrm{geo}}$ encourages the predicted embedding space to retain the pairwise relational structure of the target NVOL space.

To reduce dimensional collapse in NVOL predictions, we adopted the variance and covariance regularizers from VICReg \cite{ref36}. Let $V'$ denote centered predictions and $\sigma_j$ the standard deviation along dimension $j$. With variance floor $\gamma$,
\begin{gather}
\mathcal{L}_{\mathrm{var}} = \frac{1}{D}\sum_{j}\max\left(0, \gamma - \sigma_j\right)^2
\label{eq18}\\
\mathcal{L}_{\mathrm{cov}} = \frac{1}{D(D-1)}\sum_{i \neq j} C_{ij}^2
\label{eq19}
\end{gather}
where $C$ is the empirical covariance of centered predictions. The total prior objective combined noise prediction, point-wise, angular, pairwise geometry, and anti-collapse terms:
\begin{equation}
\begin{split}
\mathcal{L}_{\mathrm{prior}}={} & \mathcal{L}_{\mathrm{noise}}+\lambda_{\mathrm{mse}}\mathcal{L}_{\mathrm{MSE}}+\lambda_{\mathrm{cos}}\mathcal{L}_{\mathrm{cos}} \\
& \quad +\lambda_{\mathrm{geo}}\mathcal{L}_{\mathrm{geo}}+\lambda_{\mathrm{var}}\mathcal{L}_{\mathrm{var}}+\lambda_{\mathrm{cov}}\mathcal{L}_{\mathrm{cov}}
\end{split}
\label{eq20}
\end{equation}
In our implementation, we set $\lambda_{\mathrm{mse}}=0.4$ for the point-wise MSE term, $\lambda_{\mathrm{cos}}=3.0$ for the angular alignment term (the highest among all auxiliary weights, reflecting the primacy of directional accuracy for CLIP features), $\lambda_{\mathrm{geo}}=0.8$ for the pairwise geometry loss, $\lambda_{\mathrm{var}}=8\times 10^{-4}$ for the variance penalty, and $\lambda_{\mathrm{cov}}=3\times 10^{-4}$ for the covariance penalty. All weights are applied with a variance floor of $\gamma=1.0$.

\subsubsection{NVOL-to-Semantic Adapter}

The adapter $g_{\phi}$ maps layer-12 NVOL features $z^{\mathrm{NVOL}}$ to final-layer CLIP embeddings $h=g_{\phi}(z^{\mathrm{NVOL}})$, with ground-truth targets $h^{\mathrm{tgt}}$ \cite{ref14,ref17,ref28}. During training, real CLIP intermediate and final-layer features were used; at inference, NVOL features were reconstructed by the EEG-conditional diffusion prior. The adapter was trained on both domains with a domain-consistency regularizer.

We defined $g_{\phi}=g_{\phi}^{\mathrm{adapt}} \circ g^{\mathrm{proj}}$. The projection $g^{\mathrm{proj}}$ was a frozen linear map with layer normalization and L2 normalization, pretrained on EEG--CLIP alignment and applied to ViT-H/14 layer-12 features. The module $g_{\phi}^{\mathrm{adapt}}$ was a trainable bottleneck feed-forward network with layer normalization, ReLU activation, and L2 normalization \cite{ref37}. Real layer-12 features passed through both stages (ImageProjection $\rightarrow$ Adapter); diffusion-generated features, already in the target space, entered the adapter directly \cite{ref14,ref28}.

Predicted features were aligned to $h^{\mathrm{tgt}}$ with
\begin{equation}
\begin{split}
\mathcal{L}_{\mathrm{align}}={} & \lambda_{\mathrm{mse}}\mathcal{L}_{\mathrm{mse}}+\lambda_{\mathrm{smooth}}\mathcal{L}_{\mathrm{smooth}} \\
& \quad +\lambda_{\mathrm{cos}}\mathcal{L}_{\mathrm{cos}}+\lambda_{\mathrm{ctr}}\mathcal{L}_{\mathrm{ctr}}
\end{split}
\label{eq21}
\end{equation}
where $\mathcal{L}_{\mathrm{mse}}=\|h-h^{\mathrm{tgt}}\|_2^2$, $\mathcal{L}_{\mathrm{smooth}}$ is Smooth L1, $\mathcal{L}_{\mathrm{cos}}=1-\cos(h,h^{\mathrm{tgt}})$ after L2 normalization, and $\mathcal{L}_{\mathrm{ctr}}$ is a batch InfoNCE loss with temperature $\tau=0.07$ \cite{ref3,ref21,ref38}.

$\mathcal{L}_{\mathrm{align}}$ was applied separately to real and generated inputs against the same $h^{\mathrm{tgt}}$ \cite{ref14,ref39,ref40}:
\begin{equation}
\mathcal{L}_{\mathrm{align}}^{\mathrm{real}}=\mathcal{L}_{\mathrm{align}}(h_{\mathrm{real}},h^{\mathrm{tgt}}),\qquad h_{\mathrm{real}}=g_{\phi}(f_{12}^{\mathrm{real}}),
\label{eq22}
\end{equation}
\begin{equation}
\mathcal{L}_{\mathrm{align}}^{\mathrm{gen}}=\mathcal{L}_{\mathrm{align}}(h_{\mathrm{gen}},h^{\mathrm{tgt}}),
\label{eq23}
\end{equation}
where $f_{12}^{\mathrm{real}}$ denotes real layer-12 features and $h_{\mathrm{gen}}$ is obtained by passing diffusion-sampled NVOL features through the generation path. Domain consistency followed view-consistency regularization \cite{ref41}:
\begin{equation}
\mathcal{L}_{\mathrm{domain}}=1-\cos(h_{\mathrm{real}},h_{\mathrm{gen}}).
\label{eq24}
\end{equation}
The total objective was
\begin{equation}
{L}_{total}=\mathcal{L}_{\mathrm{align}}^{\mathrm{real}}+{\lambda}_{gen}\mathcal{L}_{\mathrm{align}}^{\mathrm{gen}}+{\lambda}_{domain}{L}_{domain}
\label{eq25}
\end{equation}
with weights set as: ${\lambda }_{\text{mse}}=1.0$, ${\lambda }_{\text{smooth}}=0.5$, ${\lambda }_{\text{cos}}=0.5$, ${\lambda }_{\text{ctr}}=0.2$, ${\lambda }_{\text{gen}}=1.0$, ${\lambda }_{\text{domain}}=0.1$. During training, only adapter parameters were updated, using AdamW with a cosine annealing learning-rate schedule.

\subsection{Inference pipeline}

For image synthesis, we connected the hierarchical decoding pipeline to a pretrained Stable Diffusion XL (SDXL) base model with an IP-Adapter module \cite{ref42}. Temporal--channel EEG embeddings were passed through the diffusion prior to reconstruct intermediate CLIP features and then mapped to final-layer embeddings by the semantic adapter. These decoded embeddings conditioned SDXL through IP-Adapter \cite{ref43}, which injected image-prompt information via cross-attention without fine-tuning the backbone. SDXL served as the generative backbone \cite{ref44,ref45,ref46}. At inference, no image-side CLIP features from the test stimulus were used. Generation relied on the reconstructed neural pathway alone and was intended to preserve category-level semantics together with structural cues carried by the decoded embeddings.

\subsection{Time, Spatial and Frequency analysis}

To assess the reliability of EEG-based image recognition and the sensitivity of the retrieval model across temporal, spatial, and spectral dimensions, we analyzed EEG dynamics in each domain. In the temporal domain, we averaged all training trials within each subject and computed topographic maps at 100-ms intervals. Following prior work \cite{ref47}, within-trial averaging was used to reduce masking effects associated with rapid serial visual presentation. We then compared retrieval performance across time windows, scalp regions, and frequency bands. For spatial and spectral analyses, we applied target masking by zeroing the signals of selected EEG channels or frequency bands while leaving the remaining inputs unchanged. Retrieval accuracy was recomputed on the trained model using these perturbed inputs. Channel or band importance was defined as the decrease in retrieval accuracy relative to the unmasked baseline \cite{ref48,ref49}. Results are shown in Fig.~\ref{fig5}.

\section{Results}
\label{sec:results}

\subsection{Datasets and Preprocessing}

We evaluated our approach on the THINGS-EEG dataset, which pairs EEG recordings with naturalistic images \cite{ref11}. Data were collected from 10 participants under a rapid serial visual presentation paradigm. The training set comprised 1,654 object categories, with 10 images per category and 4 presentations per image for each subject. For evaluation, we used the standard held-out test set of 200 categories; one image per category was presented to each participant across 80 trials. All experimental protocols were approved by the relevant institutional review boards, and informed consent was obtained from all participants before testing. Preprocessing followed the procedure described in \cite{ref28} to match prior work and support reproducible comparison.

\begin{figure}[!htbp]
	\centering
	\vspace{-0.5em}
	\figonesubfig{a}{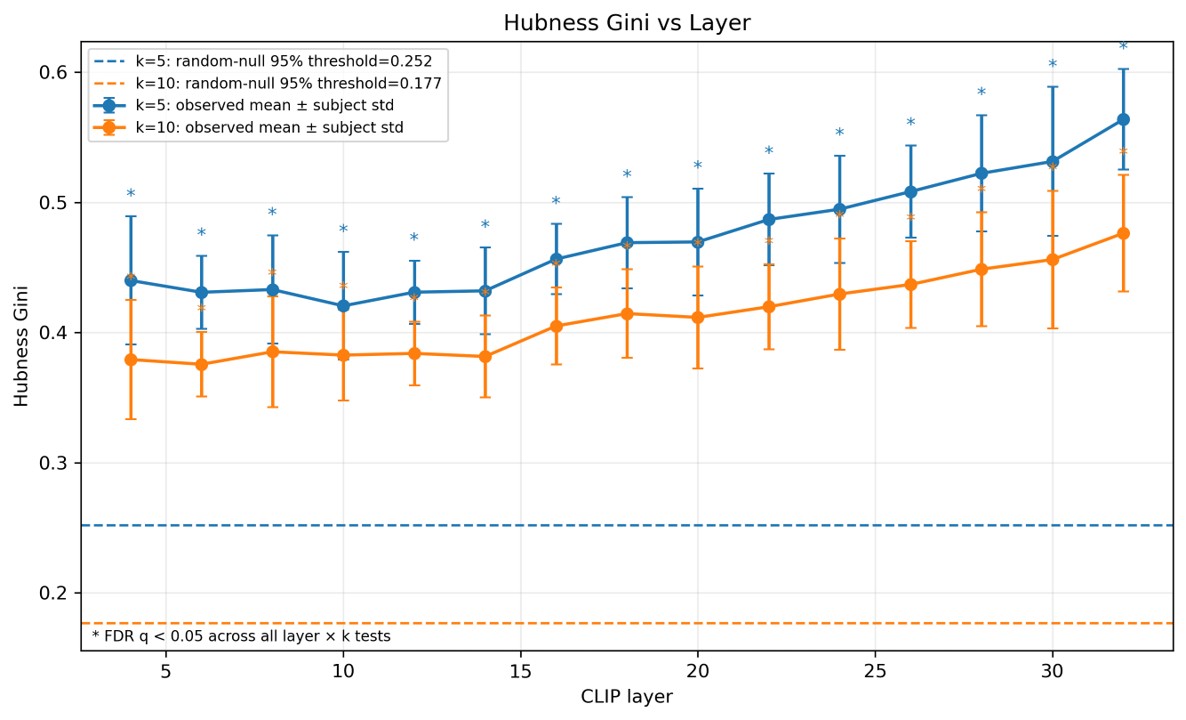}\par
	\vspace{0.05em}
	\figonesubfig{b}{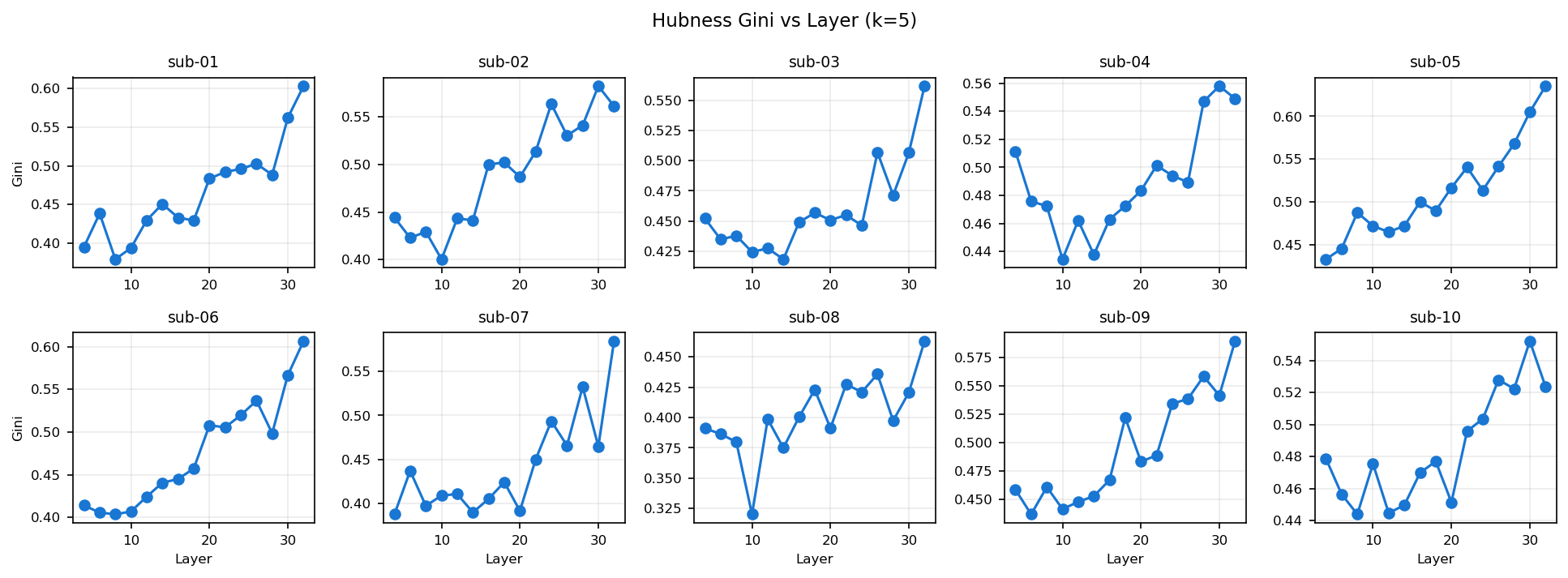}
	\caption{Hubness analysis across representation layers. (a) Variation of the hubness Gini coefficient with layer number ($k=5$, 10): the vertical axis represents the Gini coefficient, where higher values indicate more severe imbalance in nearest neighbor retrieval. (b) Variation of the Gini coefficient with layer number for each subject (sub-01 to sub-10).}
	\label{fig3}
\end{figure}

\subsection{Implementation Details}

All experiments were implemented in PyTorch and conducted on a single NVIDIA H800 GPU. We trained the model for 30 epochs using the AdamW optimizer with a learning rate of 3e-4 and a batch size of 100. For EEG encoding, ATM was employed as the neural encoder under both intra-subject and inter-subject settings. The visual encoder adopted the visual branch of CLIP ViT-H/14 \cite{ref25}, with all pre-trained weights initialized from OpenCLIP.

\subsection{Retrieval Performance}

We evaluated the image retrieval performance of our proposed method on the THINGS-EEG dataset. Comprehensive comparisons with state-of-the-art methods are reported in Tables~\ref{tab1}--\ref{tab2}.

\subsubsection{Layer-wise Retrieval Accuracy}
We varied the CLIP alignment layer from L4 to L32 and measured 200-way Top-1 and Top-5 retrieval accuracy for all ten subjects (sub-01 to sub-10). Peak accuracy occurred at intermediate layers (e.g., L10, L12, L14) in every subject (Table~\ref{tab1}). Alignment to shallow layers, which emphasize low-level features such as edges and textures, yielded lower accuracy, as did alignment to the deepest layers, which emphasize more abstract semantic content.

\subsubsection{Hubness Analysis Across Layers}

We quantified hubness, the tendency for a small subset of target images to serve as nearest neighbors for many EEG queries, using the Gini coefficient of $k$-NN retrieval statistics ($k = 5$ and $k = 10$). Each query was matched against the image feature bank. The hubness Gini index increased with CLIP layer depth (Fig.~\ref{fig3}), and at every layer it exceeded the 95th percentile thresholds derived from a random-retrieval null model ($G = 0.252$ for $k = 5$; $G = 0.177$ for $k = 10$), with all comparisons remaining significant after Benjamini--Hochberg correction ($q < 1 \times 10^{-5}$). Deep layers showed a more skewed neighbor distribution, with a few image categories occupying many top-$k$ slots. Intermediate layers had lower hubness and a more uniform neighbor distribution.

\begin{figure}[!htbp]
	\centering
	\begin{minipage}{0.65\columnwidth}
		\raggedright
		{\scriptsize\mdseries(a)}\\[0.12em]
		\includegraphics[width=\linewidth]{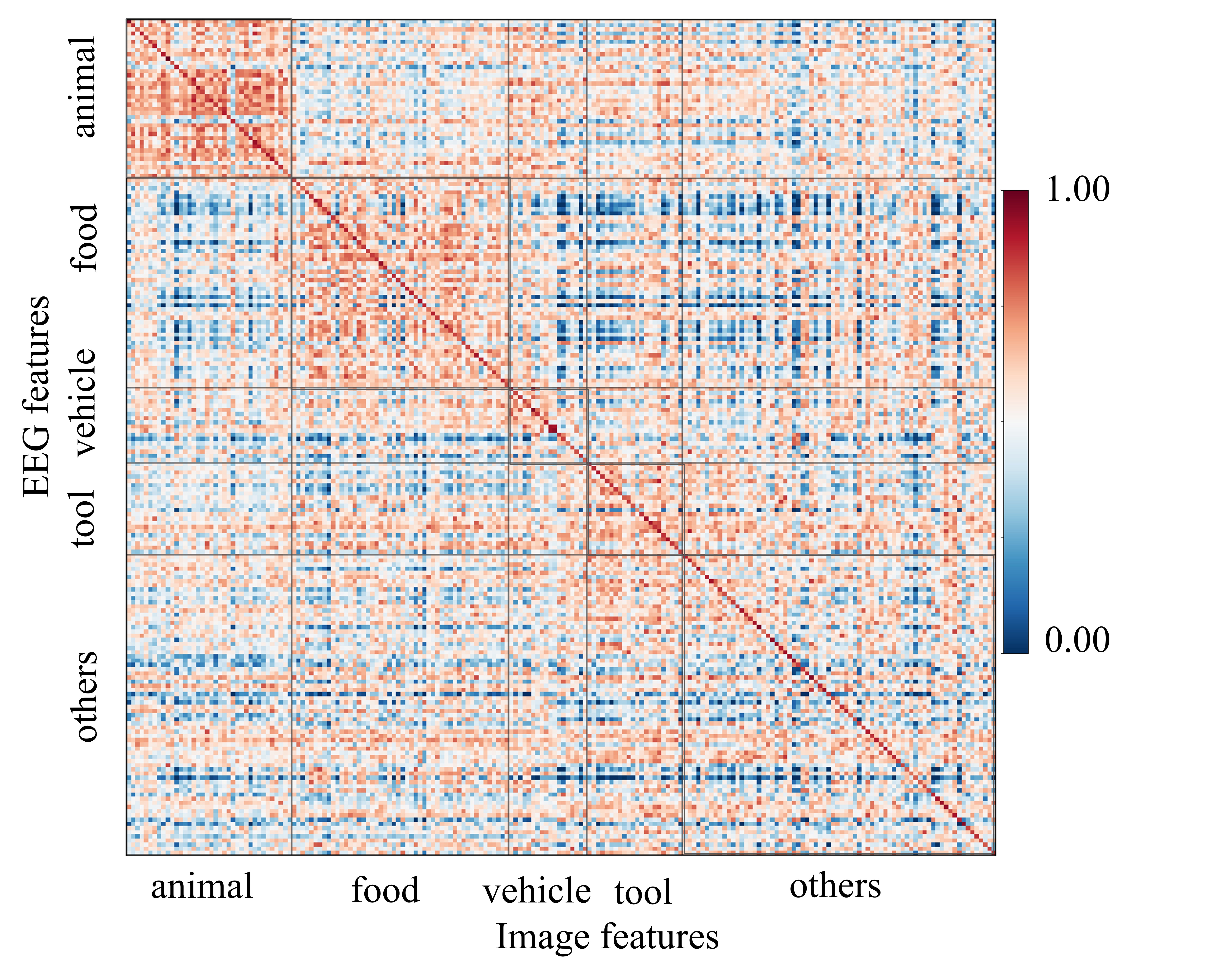}
	\end{minipage}\par
	\vspace{0.5em}
	\begin{minipage}{0.65\columnwidth}
		\raggedright
		{\scriptsize\mdseries(b)}\\[0.12em]
		\includegraphics[width=\linewidth]{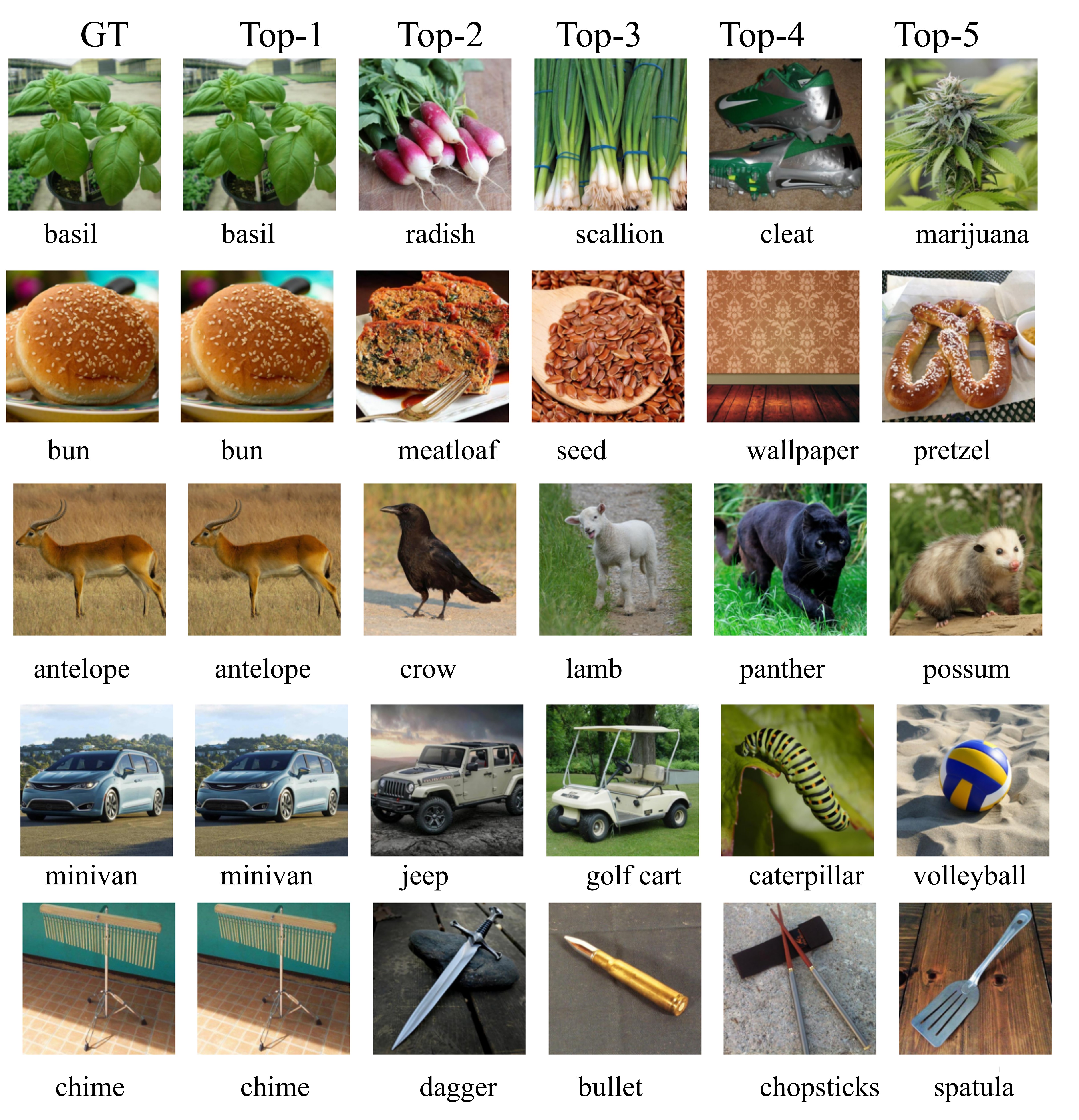}
	\end{minipage}
	\caption{Similarity analysis between EEG features and image features. (a) Representational similarity matrix of EEG for 200 concepts in the test set, with all concepts rearranged into five categories: animals, food, vehicles, tools, and others. (b) Visual comparison between ground-truth labels (first column) and top-5 predictions. GT: Ground-Truth.}
	\label{fig4}
\end{figure}
\FloatBarrier

\subsubsection{Retrieval Accuracy}

Under our best configuration (multi-layer fusion with CSLS post-processing), within-subject retrieval reached Top-1 accuracy of 86.4\% and Top-5 accuracy of 98.5\%, and between-subject retrieval reached Top-1 accuracy of 19.9\% and Top-5 accuracy of 44.1\%, exceeding prior methods on the same benchmark (Table~\ref{tab2}). Table~\ref{tab2} also reports performance for optimal single-layer retrieval (the subject-specific best single-layer accuracy selected from Table~\ref{tab1}), CSLS post-processing, multi-layer fusion alone, and the combined configuration.

\subsection{Semantic Analysis}

We computed representational similarity analysis (RSA) from cosine similarity between EEG encoder features and the corresponding image features, then normalized the values to form representational similarity matrices. For subject sub-08 (Fig.~\ref{fig4}(a)), the 200 test concepts were grouped into animals, food, vehicles, tools, and other categories and reordered accordingly. Block structure along the diagonal was consistent with category-level organization (e.g., animals, food, vehicles, tools), indicating that the intermediate layer already encodes semantically structured information beyond simple low-level visual features (color, contrast, texture). This observation suggests that the intermediate layer may serve as a sweet spot that preserves both EEG-decodable visual patterns and category-discriminative semantics. Top-5 retrieval results for sub-08 (Fig.~\ref{fig4}(b)) matched this pattern: the top-5 results for ``basil'' are mainly plants, the top-5 for ``bun'' are all food, and the top-5 for ``antelope'' are all animals.

\begin{table}[!htbp]
	\caption{Image reconstruction performance for subject-08. Metrics are divided into low-level and high-level features, where arrows indicate the direction of better performance.}
	\label{tab3}
	\centering
	\footnotesize
	\setlength{\tabcolsep}{3pt}
	\resizebox{\columnwidth}{!}{%
		\begin{tabular}{lccccc}
			\hline
			\multirow{2}{*}{Method} & \multicolumn{2}{c}{Low-level} & \multicolumn{3}{c}{High-level} \\
			\cline{2-3}\cline{4-6}
			& PixCorr $\uparrow$ & SSIM $\uparrow$ & Inception $\uparrow$ & CLIP $\uparrow$ & SwAV $\downarrow$ \\
			\hline
			CogCap(EEG)~\cite{ref51} & 0.150 & 0.347 & 0.669 & 0.715 & 0.590 \\
			ATM(EEG)~\cite{ref28} & 0.160 & 0.345 & 0.734 & 0.786 & 0.582 \\
			NEED(EEG)~\cite{ref54} & 0.162 & 0.352 & \textbf{0.742} & 0.795 & 0.575 \\
			Ours(EEG)(sub-08)(SS-Diff) & 0.165 & 0.326 & 0.710 & 0.780 & 0.571 \\
			Ours(EEG)(sub-08)(TS-Diff) & \textbf{0.176} & \textbf{0.366} & 0.737 & \textbf{0.799} & \textbf{0.552} \\
			\hline
		\end{tabular}%
	}
\end{table}
\FloatBarrier

\begin{figure}[!htbp]
	\centering
	\setlength{\figfivew}{0.98\columnwidth}%
	\begin{minipage}{\figfivew}
		\raggedright
		{\scriptsize\mdseries(a)}\\[0.12em]
		\includegraphics[width=\linewidth]{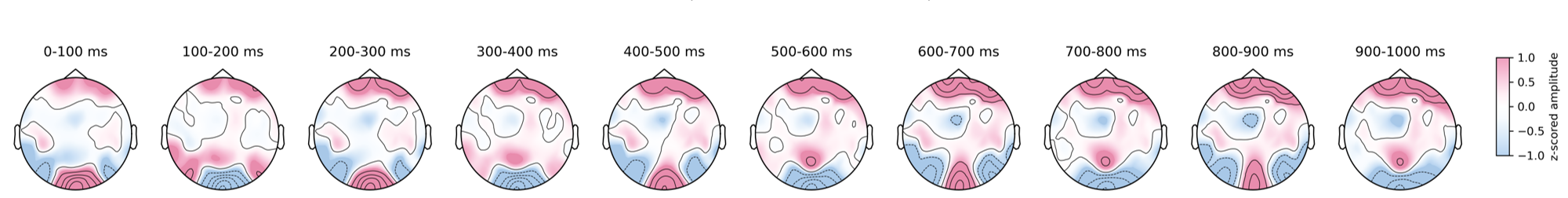}
	\end{minipage}\par\vspace{0.35em}
	\begin{minipage}{\figfivew}
		\raggedright
		{\scriptsize\mdseries(b)}\\[0.12em]
		\includegraphics[width=\linewidth]{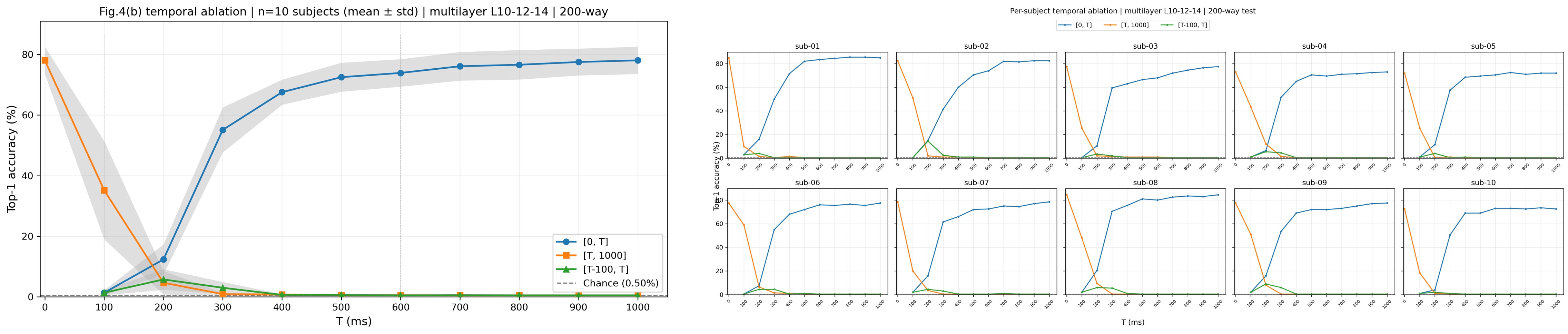}
	\end{minipage}\par\vspace{0.35em}
	\begin{minipage}[t]{0.48\figfivew}
		\raggedright
		{\scriptsize\mdseries(c)}\\[0.12em]
		\includegraphics[width=\linewidth]{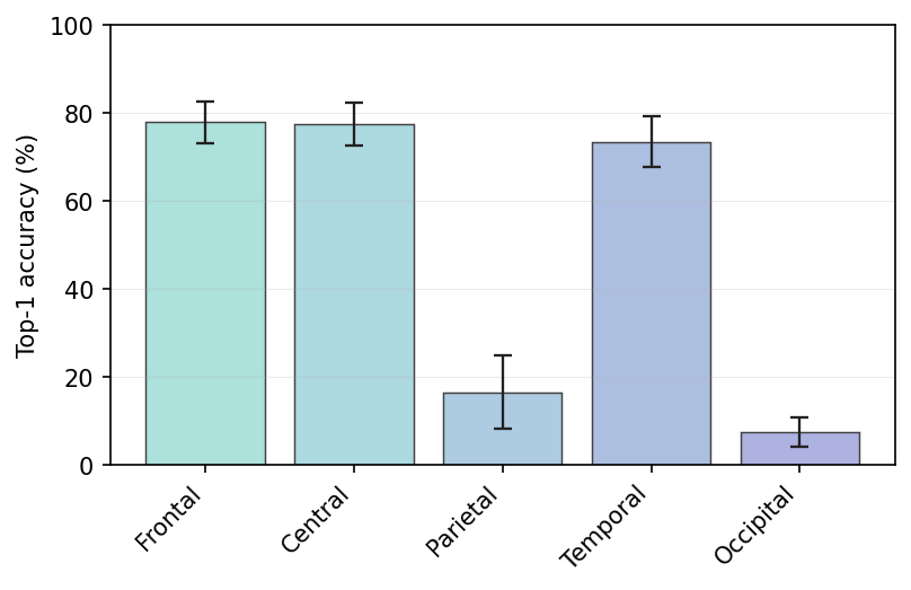}
	\end{minipage}\hspace{0.04\figfivew}%
	\begin{minipage}[t]{0.48\figfivew}
		\raggedright
		{\scriptsize\mdseries(e)}\\[0.12em]
		\includegraphics[width=\linewidth]{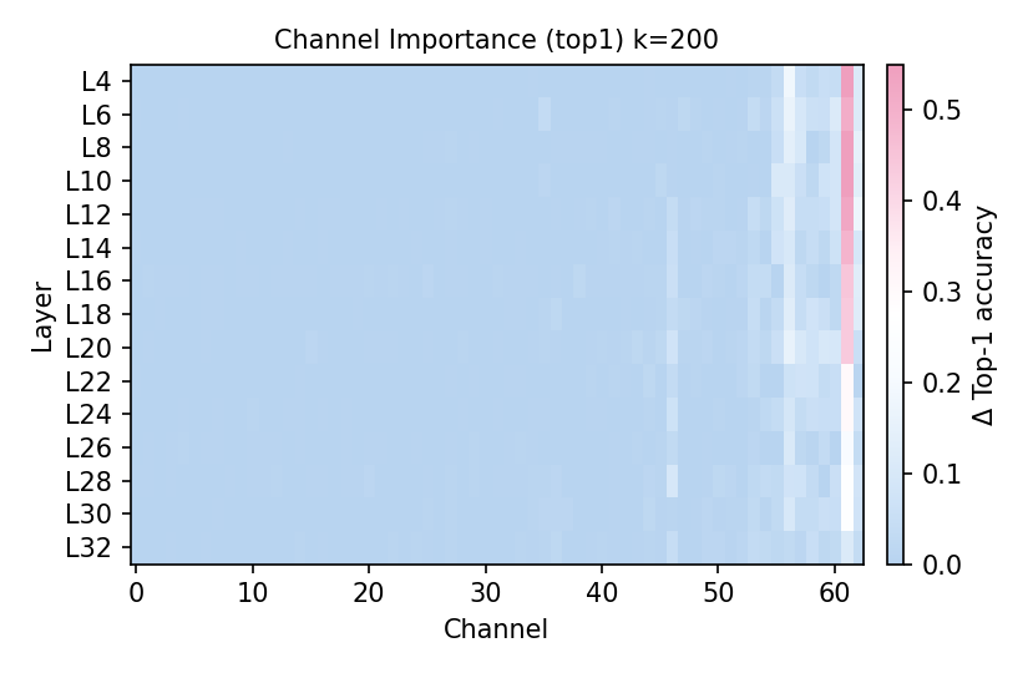}
	\end{minipage}\par\vspace{0.35em}
	\begin{minipage}[t]{0.48\figfivew}
		\raggedright
		{\scriptsize\mdseries(d)}\\[0.12em]
		\includegraphics[width=\linewidth]{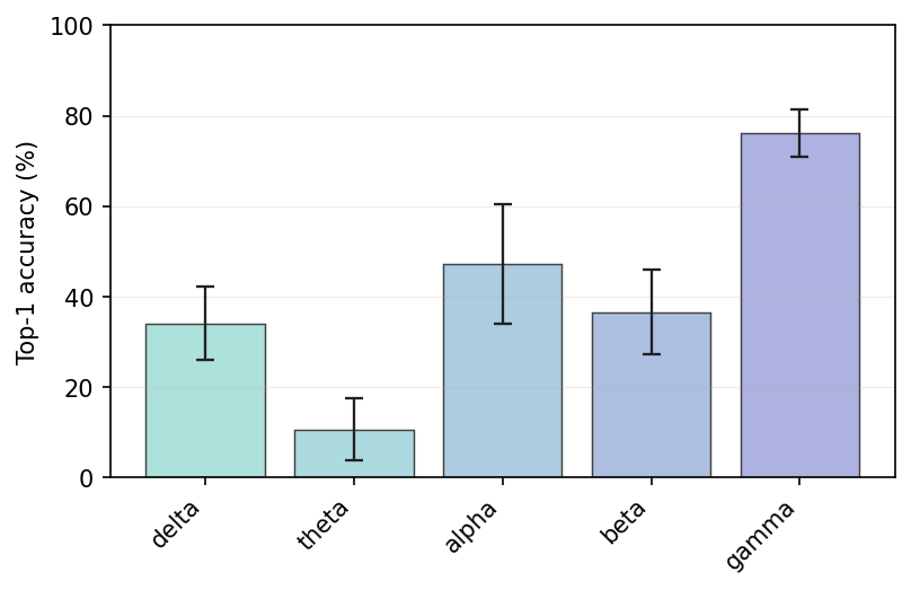}
	\end{minipage}\hspace{0.04\figfivew}%
	\begin{minipage}[t]{0.48\figfivew}
		\raggedright
		{\scriptsize\mdseries(f)}\\[0.12em]
		\includegraphics[width=\linewidth]{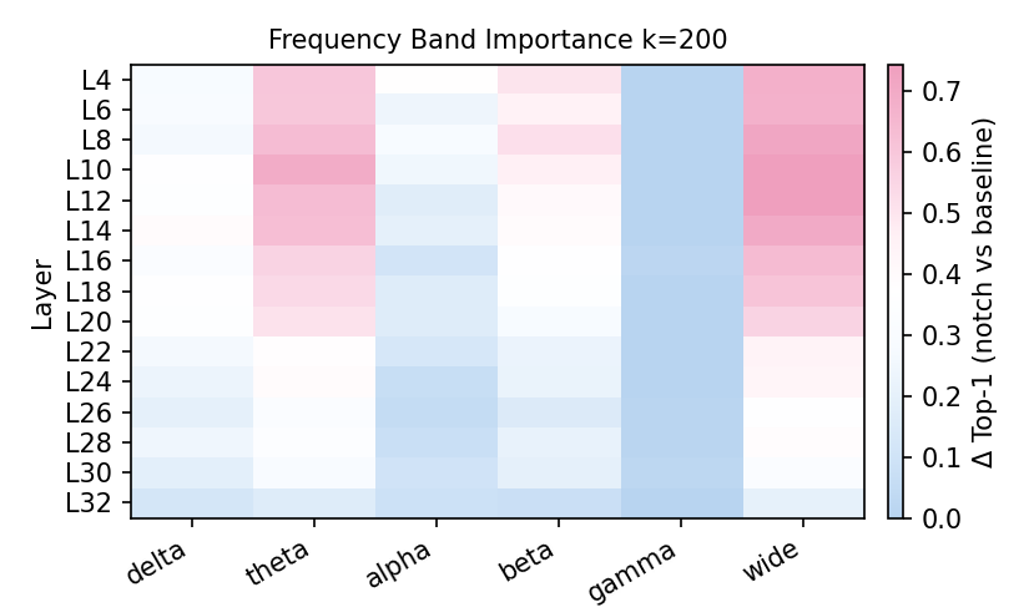}
	\end{minipage}
	\caption{Temporal, spatial, and frequency analysis of EEG. (a) Topographic maps at different time windows (every 100\,ms, averaged over all training trials of a single subject). (b) Impact of time windows on retrieval accuracy (without CSLS correction): blue, $[0,T]$; orange, $[T,1000]$\,ms; green, $[T{-}100,T]$\,ms; shaded area indicates mean$\pm$std over 10 subjects (left), with per-subject curves on the right. (c) Impact of different brain regions on retrieval performance: Top-1 retrieval rate after zeroing EEG signals in each brain region (average of 10 subjects). (d) Impact of different frequency bands on retrieval performance: Top-1 retrieval rate after removing Delta (1--4\,Hz), Theta (4--8\,Hz), Alpha (8--12\,Hz), Beta (12--30\,Hz), and Gamma (30--50\,Hz) bands (average of 10 subjects). (e) Single-channel zero-occlusion analysis: drop in Top-1 accuracy ($\Delta$Top-1) after zeroing each EEG channel, for encoders trained with different CLIP layers (200-way retrieval). Brighter colors indicate higher channel importance. The channel indices and their corresponding electrode names are listed in Table~\ref{tab4}. (f) Layer-wise band-removal analysis: drop in Top-1 accuracy ($\Delta$Top-1 $=$ baseline $-$ filtered) across CLIP layers after removing theta (4--8\,Hz), alpha (8--12\,Hz), beta (12--30\,Hz), gamma (30--50\,Hz), delta (1--4\,Hz), or broadband energy; brighter colors indicate larger performance drops.}
	\label{fig5}
\end{figure}

\subsection{Temporal, Spatial, and Spectral Analysis}
\subsubsection{EEG Temporal Analysis}
Fig.~\ref{fig5}(a) shows post-stimulus responses over occipital scalp during 0--100 ms and over temporal regions during 100--600 ms. Occipital activity also recurred at intervals consistent with the 200-ms stimulus onset asynchrony (SOA). Frontal responses increased over the epoch. We next examined how EEG time window affected retrieval performance (Fig.~\ref{fig5}(b)). Following prior work \cite{ref47}, we applied forward increment, backward decrement, and segmentation masking by zeroing samples outside each candidate window. Accuracy was highest for the [0.1, 0.6] s window. Activity in the first 0.1 s also contributed strongly: Top-1 accuracy for [0, 1] s exceeded that for [0.1, 1] s by more than 30\%.

\subsubsection{EEG Spatial Analysis}
Fig.~\ref{fig5}(c) shows the Top-1 retrieval rate for five brain regions. Masking the parietal and occipital regions led to a substantial drop in retrieval performance. Removing the temporal region also causes a certain degree of performance degradation. Channel-level analysis (Fig.~\ref{fig5}(e)) reveals a pronounced posterior-dominant topography: the Oz channel forms a persistent bright vertical band across all layers (L4--L32). Surrounding occipital (O1, O2) and parieto-occipital channels (POz, PO3, PO4, PO7, PO8) also exhibit moderate importance. In contrast, the contributions of frontal, temporal, and central channels are near baseline levels.

\begin{table}[!htbp]
	\centering
	\caption{Correspondence between channel indices and electrode names in the 10--20 montage.}
	\label{tab4}
	\scriptsize
	\setlength{\tabcolsep}{2.2pt}
	\resizebox{\columnwidth}{!}{%
	\begin{tabular}{cccccccccccccc}
		\hline
		Index & Electrode & Index & Electrode & Index & Electrode & Index & Electrode & Index & Electrode & Index & Electrode & Index & Electrode \\
		\hline
		0 & Fp1 & 9 & F3 & 18 & FC3 & 27 & C5 & 36 & TP7 & 45 & TP10 & 54 & P8 \\
		1 & Fp2 & 10 & F1 & 19 & FC1 & 28 & C3 & 37 & CP5 & 46 & P7 & 55 & PO7 \\
		2 & AF7 & 11 & F2 & 20 & FCz & 29 & C1 & 38 & CP3 & 47 & P5 & 56 & PO3 \\
		3 & AF3 & 12 & F4 & 21 & FC2 & 30 & CZ & 39 & CP1 & 48 & P3 & 57 & POZ \\
		4 & AFz & 13 & F6 & 22 & FC4 & 31 & C2 & 40 & CPz & 49 & P1 & 58 & PO4 \\
		5 & AF4 & 14 & F8 & 23 & FC6 & 32 & C4 & 41 & CP2 & 50 & Pz & 59 & PO8 \\
		6 & AF8 & 15 & FT9 & 24 & FT8 & 33 & C6 & 42 & CP4 & 51 & P2 & 60 & O1 \\
		7 & F7 & 16 & FT7 & 25 & FT10 & 34 & T8 & 43 & CP6 & 52 & P4 & 61 & OZ \\
		8 & F5 & 17 & FC5 & 26 & T7 & 35 & TP9 & 44 & TP8 & 53 & P6 & 62 & O2 \\
		\hline
	\end{tabular}%
	}
\end{table}

\subsubsection{EEG Frequency Analysis}

Fig.~\ref{fig5}(d) summarizes retrieval accuracy after band-specific filtering. Theta removal produced the largest decrease, followed by beta, delta, and alpha. Fig.~\ref{fig5}(f) plots the Top-1 accuracy reduction (baseline minus filtered) across CLIP layer depth for theta (4--8 Hz), alpha (8--12 Hz), beta (12--30 Hz), gamma (30--50 Hz), delta (1--4 Hz), and broadband. Theta showed the largest drops, peaking at intermediate layers (L8--L14). Broadband removal reduced accuracy across most layers. Gamma removal had little effect at any layer. Beta contributed modestly, with somewhat larger effects at intermediate layers. Alpha and delta effects were smaller than those of theta and broadband, and alpha effects weakened at deeper layers.

\subsection{Reconstruction Performance}

\begin{figure}[!htbp]
	\centering
	\includegraphics[width=\columnwidth]{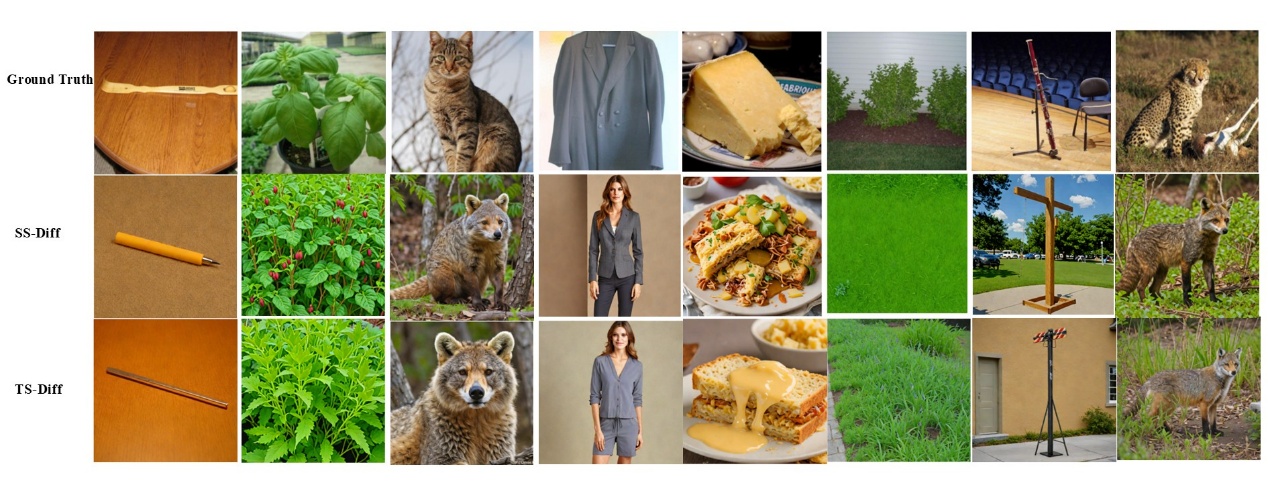}
	\caption{Comparison of generated images. Ground Truth represents the original natural images. SS-Diff (Single-Stage Final Diffusion) directly maps EEG embeddings to the final CLIP layer features via a diffusion prior, followed by image synthesis with an IP-Adapter. TS-Diff (Two-Stage Hierarchical Diffusion) first reconstructs an intermediate CLIP layer~12 feature from EEG, then passes it through a learned adapter to obtain the final semantic embedding, and finally generates the image. The two-stage approach better preserves structural alignment by leveraging a retrieval-optimal intermediate layer, leading to more faithful and coherent reconstructions.}
	\label{fig6}
\end{figure}

We evaluated image reconstruction with five metrics: PixCorr and SSIM for low-level fidelity, and Inception, CLIP, and SwAV for high-level semantic alignment. Table~\ref{tab3} compares our two-stage hierarchical diffusion model (TS-Diff) with prior EEG-based methods. TS-Diff performed best on both low- and high-level metrics, reaching PixCorr 0.176 (10.0\% above ATM) and SSIM 0.366 (5.4\% above CogCap). The single-stage variant (SS-Diff), which mapped EEG directly to the final CLIP layer, also performed competitively (SSIM 0.326). Fig.~\ref{fig6} shows representative reconstructions.

\FloatBarrier

\section{Discussion}
\label{sec:discussion}

Decoding visual signals from EEG supports both basic neuroscience and BCI design. In this study, we focused on recognizing and reconstructing natural images from EEG signals, with a core emphasis on the hierarchical compatibility between EEG representations and visual representations. Intermediate CLIP layers gave the strongest EEG--image alignment in our layer-wise screen.

We proposed a hierarchical alignment framework that aligned EEG signals with visual representations via contrastive learning. Layer-wise analyses showed that retrieval accuracy peaked at intermediate CLIP layers in all subjects, and these layers also exhibited the lowest hubness. The results showed that the visual representations at the medium level may be the most suitable for EEG-image alignment. From a neuroscientific perspective, this is consistent with the hierarchical processing theory of the ventral visual pathway. The visual representations at the medium level may be easier to decode from scalp EEG than raw pixels or highly abstract semantic categories. Our retrieval results further confirmed that by combining multi-layer interaction and the CSLS method, the Top-1 accuracy rate increased to 86.4\%, significantly superior to existing methods.

Based on the layer-wise alignment results, we developed a two-stage hierarchical reconstruction model. TS-Diff outperformed existing EEG reconstruction methods on both low- and high-level metrics, improving PixCorr by 10.0\% and SSIM by 5.4\%. The single-stage variant without an intermediate-layer prior also achieved competitive performance (SSIM $= 0.326$), indicating that EEG-to-final-layer mapping can retain meaningful visual structure even without explicit intermediate supervision. Adding the intermediate prior further improved fidelity and semantic coherence. Aligning EEG with hierarchically organized visual representations may support more effective and interpretable neural decoding.

We analyzed EEG channel, frequency, and time contributions to assess the hierarchical alignment hypothesis across spatial, spectral, and temporal domains \cite{ref55}. Image retrieval effects depended mainly on posterior scalp regions, with negligible contributions from frontal, temporal, and central areas. This pattern suggests that the EEG components driving retrieval originate from visual cortical generators rather than from frontal executive or temporal regions. The topography is compatible with early-to-mid-level visual processing and with the mid-level CLIP layer chosen for alignment, which encodes structure above low-level edges but below fully abstract semantics. We suggest that intermediate CLIP layers reflect a trade-off between EEG decodability and semantic discriminability. Deeper layers encode richer category information, but their greater abstraction from sensory input may make them harder to recover from scalp EEG. Early layers are more readily decodable, yet they provide limited category-level structure. Intermediate layers appear to retain enough visual structure for EEG decoding while already organizing representations along semantic axes. This pattern is consistent with the block-diagonal RSA structure, in which similarity matrices follow category groupings such as animals, food, and vehicles. Under this account, intermediate-layer alignment can yield both high retrieval accuracy and semantically coherent Top-5 results, without requiring EEG to access the most abstract semantic layer directly. Theta-band activity showed the strongest association with retrieval performance, and its benefit was largest at intermediate CLIP layers. Because theta has been linked to top-down attention, memory retrieval, and multimodal integration, selective theta enhancement at intermediate layers may reflect a neurophysiological correlate of mid-level visual encoding. Removing the gamma band had minimal impact, which may reflect low gamma signal-to-noise under our sampling rate, epoch length, and preprocessing pipeline, or limited coupling between scalp-recorded gamma and mid-level visual content. Temporal analyses showed that the [0.1, 0.6] s post-stimulus window carried most of the decodable information, with the first 0.1 s contributing disproportionately. This interval overlaps with prior reports of early evoked correlates of mid-level visual representations (approximately 70--200 ms) \cite{ref56,ref57,ref58}.

This study has several limitations. First, we analyzed subject-specific layer profiles but did not model cross-subject variability in alignment. Individual differences in optimal CLIP layers may reflect stable neurophysiological factors, and future work could incorporate personalized alignment or meta-learning to improve generalization across participants. Second, all experiments used THINGS-EEG. EEG decoding is sensitive to amplifier type, electrode layout, sampling rate, and task design, so performance on other datasets and acquisition setups remains to be tested. Third, it is unclear whether the hierarchical alignment framework transfers to other modalities such as MEG. Because MEG differs from EEG in source sensitivity and spatial resolution, the neural-visible optimal layer may shift across modalities; cross-dataset and cross-modal validation will be needed. Fourth, although TS-Diff improved SSIM scores, reconstructed images still lack fine texture and instance-level detail. Stronger diffusion conditioning from mid-level EEG features may further improve reconstruction fidelity.

\section{Conclusion}
\label{sec:conclusion}

In summary, through systematic retrieval, reconstruction, and spatiotemporal-spectral analysis, this study reveals the principle of hierarchical compatibility between EEG and visual representations, and demonstrates that intermediate layers are the optimal alignment anchor. This finding not only provides a new theoretical perspective for brain decoding but also lays a foundation for building more efficient and interpretable brain-computer interfaces and neural visual systems.

\section*{Acknowledgment}
The authors thank collaborators and funding agencies for their support.

\end{document}